\documentstyle[12pt,psfig]{article}

\catcode`\@=11
\def\vereq#1#2{\lower3pt\vbox{\baselineskip1.5pt \lineskip1.5pt
\ialign{$\m@th#1\hfill##\hfil$\crcr#2\crcr\sim\crcr}}}
\catcode`\@=12

\begin{document}
\begin{titlepage}
\begin{center}
August 14, 1995     \hfill    UCB-PTH-95/29\\

\vskip .25in

{\large \bf A Geometry of the Generations}\footnote{This work was
supported in part by the Director, Office of
Energy Research, Office of High Energy and Nuclear Physics, Division of
High Energy Physics of the U.S. Department of Energy under Contract
DE-AC03-76SF00098 and in part by the National Science Foundation under
grant PHY-90-21139.}

\vskip 0.3in

Lawrence J. Hall and Hitoshi Murayama

{\em Department of Physics,
     University of California\\
     Berkeley, California 94720}\\
and\\
{\em Theoretical Physics Group\\
     Ernest Orlando Lawrence Berkeley National Laboratory\\
     University of California,
     Berkeley, California 94720}

\end{center}

\vskip .3in

\vfill

\begin{abstract}
We propose a geometric theory of flavor based on  the discrete group
$(S_3)^3$, in the context of the minimal supersymmetric standard model.
The group treats three objects symmetrically, while
making fundamental distinctions between the generations.  The top quark
is the only heavy quark in the symmetry limit, and the first and second
generation squarks are degenerate.  The hierarchical nature of Yukawa
matrices is a consequence of a sequential breaking of $(S_3)^3$.
\end{abstract}

\vfill

\end{titlepage}

\renewcommand{\thepage}{\arabic{page}}
\setcounter{page}{1}

The smallness of the electroweak symmetry breaking scale and the hierarchical
nature of the Yukawa matrices 
provide two of the most important
problems of particle physics.  Weak scale supersymmetry may well play a
crucial role in the former, since it is the only symmetry which can protect
the mass of an elementary scalar. However, weak scale supersymmetry widens the
scope of flavor physics: any supersymmetric extension of the standard model
possesses eleven flavor matrices rather than the three Yukawa matrices of the
standard model. The additional eight flavor matrices all involve couplings to
squarks and sleptons, and have therefore not been directly probed
experimentally. However, rare processes, such as the $K_L$-$K_S$ mass
difference 
provide
experimental constraints on these flavor mixing matrices \cite{DG}. Hence
the problem of flavor symmetries is greatly affected by the inclusion of weak
scale supersymmetry.

It is frequently remarked that the most striking feature of the observed flavor
physics is that the top quark is the only fermion with a mass of order the weak
scale. In the context of the standard model this implies that only one entry
of the three Yukawa matrices is of order unity, while all other entries are
numerically small. 
In the context of
supersymmetric standard model,
we
find that there are now {\it two} features of flavor physics which must be
considered at the zeroth order level:
(1) the large mass of the top quark,
(2) the near absence of flavor-changing neutral currents strongly suggest
that scalars of a given charge of the light two generations are
degenerate \cite{DG,foot}. 
In this paper we explore the consequences of assuming that both of these
salient features arise from a common origin -- a flavor symmetry group
$G_f$.

The existence of an exact flavor symmetry group at high energies is very
plausible -- it is suggested by the replication of generations. However, in
many supersymmetric theories it becomes a necessity. Presumably the ultimate
theory of flavor will involve no small parameters: all the dimensionless
couplings will be of order unity and small mass ratios will result from
hierarchies of
dynamically generated mass scales, or perhaps from loop factors. If the
supersymmetry breaking squark masses appear as hard interactions in such
theories, as they do in supergravity models, then the couplings of order unity
will lead to large radiative contributions to the squark masses \cite{HKR}. The
degeneracy between
first two generation scalars
can then only be maintained if the
dimensionless couplings of the theory possess a non-Abelian flavor
symmetry $G_f$.

What should we take for $G_f$? In the context
of supergravity theories it was suggested that a U(N) invariance of the
K\"ahler
potential, where N is the total number of chiral superfields of the theory, be
used to protect the squark degeneracy \cite{HLW}.
However, in this paper we require that
$G_f$ also acts on the superpotential interactions which generate
the fermion masses, so this U(N) invariance is not possible. Flavor symmetries
which have been considered to date fall into two categories:
\par\noindent
{\bf (1) Unified} The group is such that in the symmetry
limit there is no distinction whatever between generations. This occurs
if the three generations are assigned to an
irreducible representation which has three indistinguishable
components -- such as a triplet of SU(3).
\par\noindent
{\bf (2) Asymmetric} The action of the group is such that
there is no symmetrical treatment of $N$ objects, where $N=3$ is the number of
generations. There are many examples with $G_f$ taken to be U(1)$^n$
\cite{FN} or SU(2) \cite{DKL}.

A unified $G_f$ has the advantage of providing a more complete theory of
flavor, whereas an asymmetric $G_f$ does not provide an understanding of the
difference between the generations. On the other hand, a unified $G_f$ must be
broken by couplings of order unity to obtain $m_t$, whereas an
asymmetric $G_f$, such as SU(2), can provide an understanding of the salient
flavor features even in the absence of symmetry breaking. In this paper we
propose to combine the advantages of a unified $G_f$ with those of an
asymmetric $G_f$ by introducing a third category of flavor symmetry:
\par\noindent
{\bf (3) Symmetric} The group has an action which is identical on three
objects,
yet has a representation structure which treats the generations differently.

In searching for such a group we are guided by three principles:
\par\noindent {\bf (a)}
The fields of the theory are those of the minimal supersymmetric standard
model: three generations and two Higgs doublets $H_u$ and $H_d$.
\par\noindent {\bf (b)}
The group should be a local discrete symmetry \cite{KW}.
Continous global symmetries are broken by quantum gravity
\cite{quantum} and should therefore be gauged.
However, flavor symmetries must be broken to generate Yukawa matrices,
and the breakdown of gauged flavor symmetries splits masses of different
families due to the $D$-term contribution \cite{Tahoe}.
\par\noindent {\bf (c)}
The representaton structure of the three generations should be $(1 + 2)$,
such that, in the $G_f$ symmetry limit, the top quark Yukawa coupling
is allowed, and the non-Abelian nature of the group maintains degeneracy
between the scalars of the lighter two generations.

The discrete non-Abelian group with fewest group elements
is the symmetric group $S_3$. By its
very definition it acts symmetrically on three objects. Remarkably it has two
singlets and a doublet as irreducible representations, and therefore offers
an excellent match to the flavor problem of supersymmetric theories. The action
of $S_3$ has a geometrical interpretation as all possible rotations in three
dimensions which leave an equilateral triangle invariant. The three vectors
representing the vertices of the triangle, {\boldmath{$e$}}$_1$,
{\boldmath{$e$}}$_2$ and {\boldmath{$e$}}$_3$ in Figure 1, are treated
identically by the group.
Yet the sums and differences of these vectors form a singlet
representation
({\boldmath{$v$}}$_3$) and a doublet representation ({\boldmath{$v$}}$_1$,
{\boldmath{$v$}}$_2$), whose two components have different group
properties. Despite a geometrical symmetry amongst three objects, there is also
a geometrical understanding of the differences between the generations.

The group $S_3$ has six elements:
\begin{equation}
S_3 = \{e, (12), (13), (23), (123), (132)\},
\end{equation}
where $e$ is the identity element. The two elements $(123)$ and $(132)$
are 120$^\circ$ rotation of the triangle around the axis
$\mbox{\boldmath $v$}_3 = (1, 1, 1)/\sqrt{3}$, which form {\boldmath
$Z$}$_3$ subgroup of even permutations in $S_3$.  The $(12)$, $(13)$ and
$(23)$ elements rotate the triangle by 180$^\circ$ around one of its
symmetry axes, which are odd permutations.  The vector
$\mbox{\boldmath $v$}_3$ flips its sign under odd permutations but
does not under even permutations.  This is a non-trivial
singlet representation which we call ${\bf 1}_A$, and will be identified
with the third generation later.  Two other orthogonal
vectors $\mbox{\boldmath $v$}_1 = (1, 1, -2)/\sqrt{6}$ and
$\mbox{\boldmath $v$}_2 = (-1, 1, 0)/\sqrt{2}$ form a doublet
representation ${\bf 2}$ of $S_3$.  We identify them later with first
and second generation fields, respectively.  Any {\bf 2}
representation can be written as a two-vector in $(\mbox{\boldmath
$v$}_1, \mbox{\boldmath $v$}_2)$ space.  There are only three
irreducible representations of $S_3$: ${\bf 1}_A$ and ${\bf 2}$ above
and another singlet ${\bf 1}_S$ which is a trivial
reprentation (invariant).  The ${\bf 1}_S$ representation can be obtained as
a symmetric product {\boldmath $^t x y$} of two ${\bf 2}$'s,
$\mbox{\boldmath $x$}_i$ and {\boldmath $y_i$}, while ${\bf 1}_A$ is an
anti-symmetric product $^t${\boldmath$x$}$\sigma_2${\boldmath $y$}.  The other
combinations form a ${\bf 2}$ such as $^t \mbox{\boldmath $x$} \sigma_3
\mbox{\boldmath $y$} \sim \mbox{\boldmath $v$}_1$ and
$^t \mbox{\boldmath $x$} \sigma_1 \mbox{\boldmath $y$} \sim - \mbox{\boldmath
$v$}_2$, and ${\bf 2}^3$ contains a totally symmetric invariant
$(^t \mbox{\boldmath $x$} \sigma_3 \mbox{\boldmath $y$})\mbox{\boldmath
$z$}_1 - (^t \mbox{\boldmath $x$} \sigma_1 \mbox{\boldmath
$y$})\mbox{\boldmath $z$}_2$.  Decomposition of tensor products
is shown in Table~1.

The group $S_3$ has been used before in the context of flavor physics, but from
a different perspective. The democratic ansatz for quark mass matrices
\cite{democratic}, which
leads to a heavy top quark, is known to possess the symmetry group $S_3 \times
S_3 \times SU(2) \times SU(2)$ \cite{FP}.
However, in this work the fundamental origin of
the flavor structure was assumed to come from other dynamics, perhaps BCS-like,
and $S_3$ simply appeared as an accidental consequence of this democratic
dynamics. In contrast, in this paper we argue that the supersymmetric flavor
puzzle suggests uniquely that $S_3$ is the fundamental origin of
flavor.

The work \cite{FK} is somewhat closer to our
philosophy. They advocate $G_f = Q_{2n}$,
the dicyclic dihedral groups, rather than $D_n$ or $S_n$, based on an anomaly
freedom constraint which we find to be too restrictive.  They also need
``Q-leptons" to cancel anomalies. Furthermore, although
$Q_{2n}$ possesses only singlet and doublet representations and therefore
allows a large $m_t$, this is clearly also possible with Abelian
groups. In this paper we combine the supersymmeric motivation for some
non-Abelian nature to $G_f$ with the aesthetic desire for a symmetric flavor
group.

Despite the encouraging features of $S_3$, it is not possible to satisfy
the few guiding principles (a), (b), (c) above using a single $S_3$ as
$G_f$.
For $\tilde{d}$ and $\tilde{s}$ to be
degenerate, $(d, s)_L$ and
$(d,s)_R$ should both transform as {\bf 2}. Since ${\bf 2}\times {\bf 2}
= {\bf1}_A + {\bf 1}_S + {\bf 2}$,
$m_d$ and $m_s$ are allowed by $S_3$, no matter whether
$H_d$ is assigned to ${\bf 1}_A$ or to ${\bf 1}_S$.  An enlargement of
the group is
thus necessary.  One possibility is to search for interesting structures
in larger discrete groups, such as $S_n$, $D_n$, $Q_{2n}$, and $\Delta(3
n^2)$ \cite{KS,FK}.  We find the geometric picture of the three
generations arising from the symmetric action of $S_3$ to be
sufficiently compelling that we prefer to replicate $S_3$ factors. Hence
we consider a group $S^{Q}_3
\times S^{U}_3 \times S^{D}_3$
with each of $Q$, $U$, $D$ transform as ${\bf 1}+ {\bf 2}$ under its own
$S_3$, while  transforming trivially under other factors.

We identify the third generation with ${\bf 1}_A$ rather than ${\bf
1}_S$, because
we would like to consider the discrete flavor group as an anomaly-free gauge
symmetry.
The only anomaly one can discuss with the low-energy particle content alone is
$S_3 \times H^2$ where $H$ is either SU(2) or SU(3) in the standard
model \cite{DB}.  Consider the element
$(12)$, which leaves ${\bf 1}_S$ and {\boldmath$v$}$_1$ in {\bf 2}
invariant but changes sign of ${\bf 1}_A$ and {\boldmath$v$}$_2$ in {\bf 2}.
To avoid an anomaly, the total number of ${\bf 1}_A$ and ${\bf 2}$ with a given
quantum number has to be even.  In our context, this requirement
uniquely selects ${\bf 1}_A + {\bf 2}$.  The anomaly freedom
of this choice can be easily understood by noting a vector ${\bf 3}$ in
an anomaly free group SO(3) decomposes to ${\bf 1}_A + {\bf 2}$.
Furthermore, this choice is precisely the one which allows a geometric
interpretation of families in terms of rotations.  It is interesting to
see that the three generations, although in a reducible
representation ${\bf 1}_A + {\bf 2}$, require each other to render the
theory consistent quantum mechanically.

The flavor transformation properties of the quarks are shown in Table 2.
The quantum number of $H_u$ is fixed to allow $m_t$ by $G_f$.
Anomaly freedom
then dictates an identical transformation for $H_d$.
Because of this charge assignment, only the top quark is heavy in the
$(S_3)^3$ symmetric
limit.  The top-bottom asymmetry, or more generally the up-down
asymmetry, is built into the representation structure of the Higgs.  On the
other hand, squark mass matrices all have the form $\mbox{diag}(M_1^2,
M_1^2, M_3^2)$.
The lepton sector will be discussed elsewhere.

Now we consider the breaking of $(S_3)^3$ symmetry and discuss its
consequence on the Yukawa and squark mass matrices.
In order to keep the number of breaking parameters as small as possible,
we take the following ``minimal'' form of the Yukawa matrices \cite{CM},
\begin{eqnarray}
& & Y_u = \left( \begin{array}{cc|c}
	h_u & O(\sqrt{h_u h_c}) & - h_t \lambda^3 A (\rho + i \eta)\\
	O(\sqrt{h_u h_c}) & h_c & - h_t \lambda^2 A \\ \hline
	0 & 0 & h_t
		\end{array} \right), \\
& & Y_d = \left( \begin{array}{cc|c}
	h_d & h_s \lambda & 0\\
	O(h_s \lambda) & h_s & 0 \\ \hline 0 & 0 & h_b
		\end{array} \right),
\end{eqnarray}
which correctly reproduce Cabbibo--Kobayashi--Maskawa (CKM) matrix in
Wolfenstein parametrization.  The quark masses are related to the Yukawa
couplings by $m_{u,c,t} = h_{u,c,t} \langle H_u \rangle$ and
$m_{d,s,b} = h_{d,s,b} \langle H_d \rangle$.
We assumed $(Y_u)_{12}$ and $(Y_u)_{21}$ to be $O(\sqrt{h_u h_c})$,
because larger off-diagonal elements need a fine-tuning in the
determinant.  We actually do not need these elements and can set them
vanishing, but we kept them to make the discussion more general.  The
same comment applies to $(Y_d)_{21}$.
The Cabbibo angle originates in the down sector and it may be
possible to keep the famous relation $\lambda \sim \sqrt{m_d/m_s}$.

The largest breaking parameters in the Yukawa matrices are $h_b$ which
transforms as $({\bf 1}_S, {\bf 1}_A, {\bf 1}_A)$ and $h_t \lambda^2 A$
as $({\bf 2}, {\bf 1}_S, {\bf 1}_S)$.  $h_b$ breaks $S_3^U \times S_3^D$
down to a subgroup $S_3^U \times S_3^D / \mbox{\boldmath$Z$}_2$, where
(even,~odd) and (odd,~even) elements are removed.  Note that the
diagonal subgroup $S_3^{U,D}$ is a subgroup of the unbroken symmetry.
$h_t \lambda^2 A$ is a {\boldmath$v$}$_1$ element in a doublet, and
breaks $S_3^Q$ to $S_2^Q \simeq \mbox{\boldmath$Z$}_2 = \{e, (12)\}$.
This {\boldmath $Z$}$_2$ flips the sign of second generation and Higgs
fields, while leaving first generation field unchanged.  Therefore $Q_2$
can acquire a Yukawa coupling while $Q_1$
cannot.  $h_c$ and $h_s$ belong to breaking parameters $({\bf
2}, {\bf 2}, {\bf 1}_S)$ and $({\bf 2}, {\bf 1}_A, {\bf 2})$,
respectively, and break the diagonal $S_3^{U,D}$ to {\boldmath$Z$}$_2$
as well, which still keep all first generation fields massless.  After
including the smaller breaking parameters, the symmetry $(S_3)^3$ is
completely broken.  In this way, the hierarchical pattern of the Yukawa
matrices can be understood as a sequential breaking of the flavor
symmetry.

Now we turn to the squark mass matrices.  Since the constraints from the
flavor-changing neutral currents are at best of order a few times
$10^{-3}$, we work out the non-degeneracy in squark masses
down to this order.  It is straight-forward to work out how the
breaking parameters enter the scalar matrices.  For $m^2_Q$ matrix, the
leading correction comes from $({\bf 2}, {\bf 1}_S, {\bf 1}_S)$ with
{\boldmath $v$}$_1$ and {\boldmath $v$}$_2$ components of $O(h_t \lambda^2
A)$ and $O(h_t \lambda^3 A (\rho + i\eta))$, respectively.  Therefore,
\begin{eqnarray}
m^2_{Q} \sim
	\left( \begin{array}{cc|c}
	M_1^2 + m^2 h_t \lambda^2 A & m^2 h_t \lambda^3 A (\rho + i\eta) &
		- m^{\prime2} h_t \lambda^2 A\\
	m^2 h_t \lambda^3 A (\rho + i\eta) & M_1^2 - m^2 h_t \lambda^2 A &
		m^{\prime2} h_t \lambda^3 A (\rho + i\eta) \\
	\hline
	- m^{\prime2} h_t \lambda^2 A &
		m^{\prime2} h_t \lambda^3 A (\rho - i\eta)  & M_3^2
                \end{array} \right),
\end{eqnarray}
where a possible correction to $(m_Q^2)_{33}$ was absorbed into
$M_3^2$.
Here and
hereafter, $m^2$ and $m^{\prime2}$ are arbitrary numbers comparable to
$M_1^2$ and $M_3^2$, and they are in general different for $Q$, $U$,
$D$.
For the $m_U^2$ matrix, the only correction comes from the square of $({\bf
2}, {\bf 2}, {\bf 1}_S)$ breaking parameter of $O(h_c^2)$.  The resulting
form is \begin{equation}
m^2_{U} \sim
	\left( \begin{array}{cc|c}
	M_1^2 + m^2 h_c^2 & m^2 \sqrt{h_c^3 h_u} &
		- m^{\prime2} h_c^2 \\
	m^2 \sqrt{h_c^3 h_u}& M_1^2 - m^2 h_c^2 &
		m^{\prime2} \sqrt{h_c^3 h_u} \\
	\hline
	- m^{\prime2} h_c^2 &
		m^{\prime2} \sqrt{h_c^3 h_u}  & M_3^2
                \end{array} \right).
\end{equation}
The $m_D^2$ matrix receives corrections from two sources at the leading
order.  One is the square of the
$({\bf 2}, {\bf 1}_A, {\bf 2})$ breaking parameter of $O(h_s^2)$,
and the other is a product of three breaking parameters
$({\bf 1}_S, {\bf 1}_A, {\bf 1}_A)$, $({\bf 2}, {\bf 1}_A, {\bf 2})$,
and $({\bf 2}, {\bf 1}_S, {\bf 1}_S)$ of $O(h_s h_b h_t A \lambda^2)$.  They
are of the same order of magnitude and have the same group theoretical
structure $({\bf 1}_S, {\bf 1}_S, {\bf 2})$.  We keep only the first
for simplicity and obtain
\begin{equation}
m^2_{D} \sim
	\left( \begin{array}{cc|c}
        M_1^2 m^2 h_s^2 & m^2 h_s^2 \lambda &
                - m^{\prime2} h_s^2 \\
        m^2 h_s^2 \lambda & M_1^2 - m^2 h_s^2 &
                m^{\prime2} h_s^2 \lambda \\
        \hline
        - m^{\prime2} h_s^2 &
                m^{\prime2} h_s^2 \lambda  & M_3^2
                \end{array} \right).
\end{equation}

The authors of
\cite{NS} listed the constraints on the off-diagonal mass matrix
elements for $m_{\tilde{q}} \sim 1$~TeV in the basis where the Yukawa
matrices are diagonal.  We adopt their notation and list the
constraints in Tables 2, 3, 4.  It is clear that our mass matrices
satisfy all constraints rather easily.  We have not discussed the
left-right mixing mass matrix so far, but they are tightly constrained
by the $S_3^3$ symmetry as well.  The breaking parameters enter the
mixing mass matrix in the same manner as in the Yukawa matrices.  It is
easy to work them out and see that the constraints are easily satisfied.

A natural question is how much stronger the constraints become when we
introduce further breaking parameters and introduce mixing in the
right-handed fields as well.  The off-diagonal elements of $m_D^2$ and
$m_U^2$ can be much larger than the above estimates.  However, they are
at most of the same order as those in $m_Q^2$ if we assume a
similar order of mixing angles in the right-handed fields.
On the other hand, constraints become even weaker if we attribute all
CKM angles to the down sector, since the breaking parameters are then
proportional to $h_b$ rather than $h_t$.  A potentially dangerous
breaking is that in $({\bf 1}_S, {\bf 1}_S, {\bf 1}_A)$ or $({\bf 1}_A,
{\bf 1}_S, {\bf 1}_S)$, which do not contribute to the Yukawa matrices.
However they are presumably as small as $h_u$ or $h_d$ because they
break the {\boldmath$Z$}$_2$ symmetry which keeps the first generation
fields massless.

In summary, we proposed a geometric theory of flavor based on the
discrete group $(S_3)^3$.  The group acts symmetrically on three objects, yet
gives fundamentally different characteristics to each generation.
The three generations belong to a reducible representation
${\bf 2}+{\bf 1}_A$; although they are not unified, they require
each other for anomaly cancellations.
Only the top quark is heavy in the symmetry limit, and
first- and second-generation squarks are degenerate.  Hierarchical
Yukawa matrices can be understood as a consequence of sequential
symmetry breaking.  Flavor-changing processes are highly suppressed,
allowing squarks at Tevatron energies.

\begin{table}
\begin{center}
\begin{tabular}{c|ccc}
$\otimes$ & ${\bf 1}_S$ & ${\bf 1}_A$ & ${\bf 2}$\\ \hline
${\bf 1}_S$ & ${\bf 1}_S$ & ${\bf 1}_A$ & ${\bf 2}$\\
${\bf 1}_A$ & ${\bf 1}_A$ & ${\bf 1}_S$ & ${\bf 2}$\\
${\bf 2}$ & ${\bf 2}$ & ${\bf 2}$ & ${\bf 1}_A \oplus
{\bf 1}_S \oplus {\bf 2}$
\end{tabular}
\end{center}
\caption[tensor]{Decomposition of tensor product of two
representations into irreducible representations.}
\end{table}

\begin{table}
\begin{center}
\begin{tabular}{|c|ccc|cc|}
& $Q$&$U$&$D$&$H_u$&$H_d$\\
\hline
$S^Q_3$&$({\bf 1}_A, {\bf 2})$&-&-&${\bf 1}_A$&${\bf 1}_A$\\
$S^U_3$&-&$({\bf 1}_A, {\bf 2})$&-&${\bf 1}_A$&${\bf 1}_A$\\
$S^D_3$&-&-&$({\bf 1}_A, {\bf 2})$&-&-
\end{tabular}
\end{center}
\caption[charge]{Quantum number assignments of the fields under
$(S_3)^3$ symmetry.  $Q$ refers to left-handed quark doublets, $U$ ($D$)
to right-handed up(down)-type quarks.}
\end{table}

\begin{table}
\begin{center}
\begin{tabular}{c|cccc}
& $(\delta^d_{LL})_{12}$ & $(\delta^d_{RR})_{12}$ &
$(\delta^d_{LR})_{12}$ &
$\langle \delta^d_{12}\rangle$ \\ \hline
{upper bound \cite{NS}} & 0.05 & 0.05 & 0.008 & 0.006 \\
{this model} & $h_t A \lambda^3$ 
	& $h_s^2 \lambda^2$ & $h_s^2 \lambda$
	& $\sqrt{(\delta^d_{LL})_{12} (\delta^d_{RR})_{12}} $
\end{tabular}
\end{center}
\caption[12d]{The constraints and the consequence of $(S_3)^3$ symmetry
on the mass splittings in $\tilde{d}$-$\tilde{s}$.}
\end{table}

\begin{table}
\begin{center}
\begin{tabular}{c|cccc}
& $(\delta^d_{LL})_{13}$ & $(\delta^d_{RR})_{13}$ &
$(\delta^d_{LR})_{13}$ &
$\langle \delta^d_{13}\rangle$\\ \hline
{upper bound \cite{NS}} & 0.1 & 0.1 & 0.06 & 0.04 \\
{this model} & $h_t A \lambda^2$ & $h_s^2$ & $h_b h_t A \lambda^3$ &
	$\sqrt{(\delta^d_{LL})_{13} (\delta^d_{RR})_{13}}$
\end{tabular}
\end{center}
\caption[12d]{The constraints and the consequence of $(S_3)^3$ symmetry
on the mass splittings in $\tilde{d}$-$\tilde{b}$.}
\end{table}

\begin{table}
\begin{center}
\begin{tabular}{c|cccc}
& $(\delta^u_{LL})_{12}$ & $(\delta^u_{RR})_{12}$ &
$(\delta^u_{LR})_{12}$ &
$(\delta^u_{LR})_{12}$ \\ \hline
{upper bound \cite{NS}} & 0.1 & 0.1 & 0.06 & 0.04 \\
{this model} & $h_t A \lambda^3$ 
	& $\sqrt{h_u h_c^3}$ &
	$\sqrt{h_u h_c^3}$ &
	$\sqrt{(\delta^u_{LL})_{12} (\delta^u_{RR})_{12}}$
\end{tabular}
\end{center}
\caption[12d]{The constraints and the consequence of $(S_3)^3$ symmetry
on the mass splittings in $\tilde{u}$-$\tilde{c}$.  We assumed that the
rotation angle between $u$ and $c$ is $O(\sqrt{h_u/h_c})$.}
\end{table}

\begin{figure}
\centerline{
\psfig{file=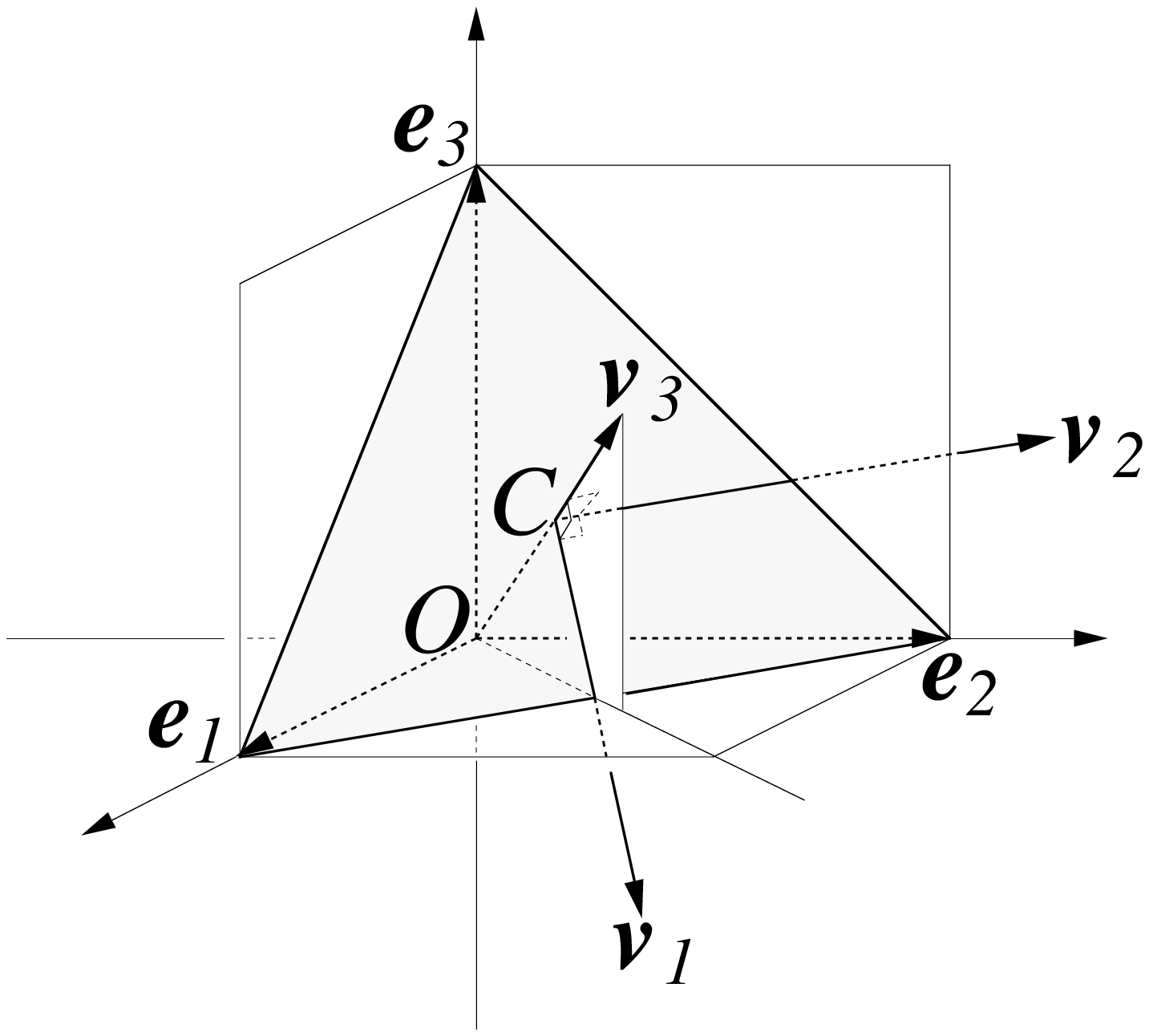,width=6cm}
}
\caption[Fig1]{$S_3$ acts as a rotation of the triangle spanned by three
orthonomal vectors {\boldmath$e$}$_{1,2,3}$.  The vector
{\boldmath$v$}$_3$ corresponds to the ${\bf 1}_A$ representation, and
two vectors {\boldmath$v$}$_{1,2}$, in the plane of the triangle, to the
${\bf 2}$ representation.}
\end{figure}


\begin{thebibliography}{99}
\bibitem{foot} This degeneracy is not a necessity, as flavor-changing
effects in the down sector can be greatly suppressed by assuming that
the origin of the CKM matrix resides in the up
sector \cite{NS}.  However, it is somewhat contrived to choose the
flavor symmetries of the theory so that the flavor violations arise only
in the sector of the theory where the experimental constraints are the
weakest.
\end{thebibliography}
\end{document}